\documentclass[runningheads]{llncs}
\usepackage{graphicx}
\usepackage{makecell}
\begin{document}
\title{Potential of deep features for opinion-unaware, distortion-unaware, no-reference image quality assessment\thanks{Supported by NSERC Discovery Grant and DND Supplement.}}
%
%
\author{Subhayan Mukherjee\inst{1}\orcidID{0000-0002-6479-3893} \and
Giuseppe Valenzise\inst{2} \and
Irene Cheng\inst{1}\orcidID{0000-0001-9699-4895}}
\authorrunning{S. Mukherjee et al.}
%
\institute{University of Alberta, Edmonton AB T6G 2R3, Canada
\email{\{mukherje,locheng\}@ualberta.ca} \and
CNRS - CentraleSupelec - Université Paris-Sud, 91192 Gif-sur-Yvette Cedex, France
\email{giuseppe.valenzise@l2s.centralesupelec.fr}}
\maketitle              
\begin{abstract}
Image Quality Assessment algorithms predict a quality score for a pristine or distorted input image, such that it correlates with human opinion. Traditional methods required a non-distorted ``reference'' version of the input image to compare with, in order to predict this score. However, recent ``No-reference'' methods circumvent this requirement by modelling the distribution of clean image features, thereby making them more suitable for practical use. However, majority of such methods either use hand-crafted features or require training on human opinion scores (supervised learning), which are difficult to obtain and standardise. We explore the possibility of using deep features instead, particularly, the encoded (bottleneck) feature maps of a Convolutional Autoencoder neural network architecture. Also, we do not train the network on subjective scores (unsupervised learning). The primary requirements for an IQA method are monotonic increase in predicted scores with increasing degree of input image distortion, and consistent ranking of images with the same distortion type and content, but different distortion levels. Quantitative experiments using the Pearson, Kendall and Spearman correlation scores on a diverse set of images show that our proposed method meets the above requirements better than the state-of-art method (which uses hand-crafted features) for three types of distortions: blurring, noise and compression artefacts. This demonstrates the potential for future research in this relatively unexplored sub-area within IQA.

\keywords{Image Quality Assessment \and Opinion Unaware \and Distortion Unaware \and No Reference \and Deep Learning.}
\end{abstract}
\section{Introduction}
Before the invention of digital cameras and other consumer grade digital imaging devices, the capture of images was quite limited. The time from capture to visualization was significant as in case of film cameras, we had to develop the photos in a dark room with chemical solutions. However, nowadays, with the advent of digital photography, coupled with the explosion in bandwidth of transmission channels like the Internet and social media, a tremendous volume of images are being captured and shared. Curating this huge volume of visual data that is being captured, stored, transmitted and viewed is a challenging task. Transmission of visual content occupies a large amount of Internet bandwidth. To meet the in-time transmission constraints limited by hardware resources, images and videos are usually processed and compressed before transmission and storage. Quality reductions happen as a trade-off between limited hardware resources and visual fidelity. Thus, automatic quality assessment methods are desirable to estimate the human-perceived quality measure, to replace subjective human perception. The distortions can be wide and varied; A common type of distortion is noise. Noise affects images captured using all types of sensors (optical or otherwise) and can even get injected into the image signal during transmission, for example through a telecommunications channel like television. The image quality research domain has focused on quality assessment of natural images and videos, since that is the dominant form of images we deal with everyday.

\begin{figure}
\centering
\includegraphics[width=80mm]{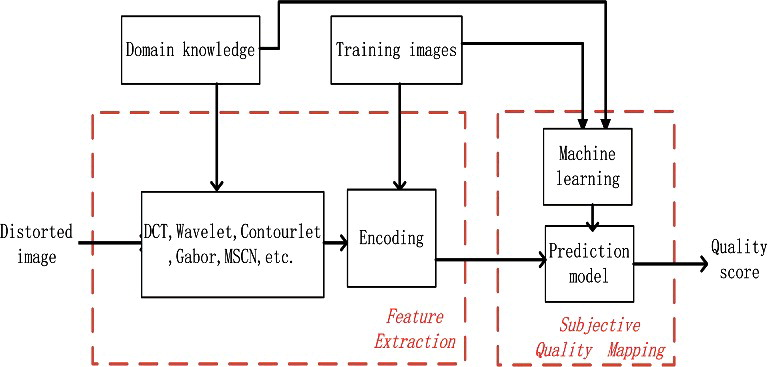}
\caption{General framework of NR-IQA algorithms.}
\label{fig:nriqa}
\end{figure}

Existing IQA methods can be classified into three categories, based on the amount of the information that is available to the method: Full-Reference (FR), Reduced-Reference (RR), and No-Reference (NR) methods \cite{Mittal2015}. FR quality assessment requires both access to the distorted images or videos as well as the clean references. RR utilizes the limited information, depending on the actual situation, regarding the reference, rather than the full reference itself, together with the distorted images. NR approaches usually perform automatic quality assessment of the images or videos using only the distorted sources. We focus specifically on the No-Reference quality assessment of images. We can draw a parallel with the Human Visual System (HVS) which has the ability to distinguish between natural and distorted scenes based on few visual memories learned while the human brain processes visual information in various ways. Most NR-IQA algorithms follow the two step process (1) feature extraction and (2) quality prediction \cite{Xu2016}. The schematic diagram of the same is shown in Fig.\ref{fig:nriqa}.

Recent attempts at image quality assessment aim to mimic the response of the HVS which can mask certain artefacts depending on the location of the artefact in the image and the surrounding image content, brightness, contrast, etc. Such ``perceptual'' methods can be categorized into different classes based on availability of subjective rating scores for the distorted images, and knowledge about the type of possible distortions. Of those, we are interested in the specific case where the subjective rating scores are unavailable (opinion-unaware/OU) and the type of distortions is unknown (distortion-unaware or `general'). Convolutional Neural Networks (CNNs) have been successfully used in different computer vision tasks, including style transfer, image generation, etc. Internal activations of deep convolutional networks, trained on high-level image classification tasks, have been found to be useful for natural feature representation, and are used to mimic human perception. In this work, we propose perceptual quality assessment using such ``deep'' features, and objectively evaluate its effectiveness.

\subsection{Summary of Contributions}
\label{sec:sumcon}
IQA has already been a quite active research area for several decades; Thus, our contribution is focused to a very specific sub-area (which has not received much attention) at the intersection of the following paradigms:

\begin{enumerate}
    \item Using deep instead of hand-crafted features for image representation.
    \item Using non-parametric instead of parametric approach to fit the distribution of pristine image features, and compare it to that of query image features.
    \item Using unsupervised instead of supervised learning, thus removing the dependence on manual labelling (subjective scores) to train the IQA model.
\end{enumerate}

The rest of the paper is organized as follows: Section \ref{sec:background} summarizes related IQA research. Section \ref{sec:method} describes how proposed method design differs from existing ones. Section \ref{sec:results} presents experimental results. Section \ref{sec:confut} concludes the paper.

\section{Background and Related Work}
\label{sec:background}

IQA methods may be broadly classified as reference or no-reference (``NR/blind'') \cite{Mittal2015}. In simple terms, all the reference-based methods try to estimate some form of \textit{distance} between the reference (``clean'') image and the input/query image. The larger the distance, the greater the distortion score. No-reference methods predict the image quality based on the distorted image itself, without the need of a reference image. Most NR-IQA algorithms follow the two step process (1) feature extraction and (2) quality prediction \cite{Xu2016}. NR-IQA algorithms are further categorized as (a) distortion-specific and (b) general / universal. The former assumes that the distortion-type is known, and employ distortion model(s) to predict one or more types of distortions in the image like noise, blur, blocking, ringing etc. to estimate its overall visual quality. The latter broadly assumes that natural scenes contain repeating patterns with a definite set of statistical properties called Natural Scene Statistics (NSS) in the spatial \cite{6272356} or transformed \cite{5430991} domain, and distortions to natural images distort these properties in measurable ways. Researchers also explored more effective ways to characterize structural and contrast distortions by modelling the gradient magnitudes of natural images as Weibull distribution, in the works popularized as (IL)NIQE \cite{6353522,7094273}. 

For score prediction, a popular approach is to fit the joint distribution of the feature vector and the associated opinion scores to a subset of the training data \cite{5430991}. In this case, the score prediction amounts to maximizing the probability of opinion score of test data, given the test data feature vector. Other approaches quantify the distance in sparse feature space between reference and distorted images \cite{7032268} in a manner that is both opinion-unaware and distortion-unaware. More recent works are based on machine learning, and extend to High Dynamic Range (HDR) images \cite{8123879}, though for this method the training is opinion-aware.

Very recent methods in the opinion/distortion unaware domain use (IL)NIQE features, but consider activations in pre-trained deep neural networks to select salient patches. They assign more weight to scores from those patches over others during score aggregation \cite{Zhang2018}. (IL)NIQE features can also reliably predict quality of multi-spectral images \cite{8672130}. Few methods even tread the boundaries of opinion (un-)awareness and/or (no-)reference \cite{Pan_2018_CVPR,Lin_2018_CVPR}. But even they do not operate at the intersection of paradigms outlined in Section \ref{sec:sumcon}, which motivates our research.

\section{Proposed Method}
\label{sec:method}

\begin{figure}
\centering
\includegraphics[width=120mm]{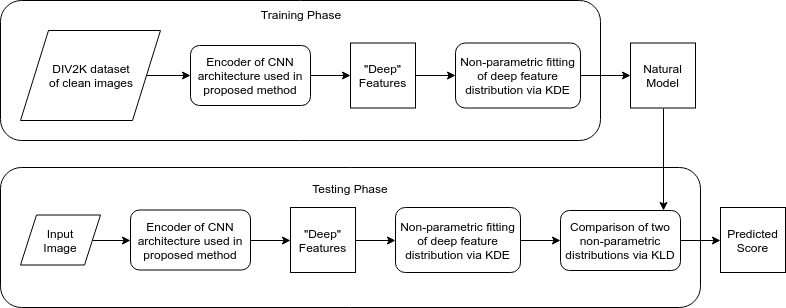}
\caption{Architecture of proposed method.}
\label{fig:proposed_arch}
\end{figure}

Below, we briefly outline how the proposed method's functionality as visualised in Fig. \ref{fig:proposed_arch} conforms to the intersection of paradigms outlined in Section \ref{sec:sumcon}:

\subsection{Deep features for image representation}
\begin{figure}
\centering
\includegraphics[width=80mm]{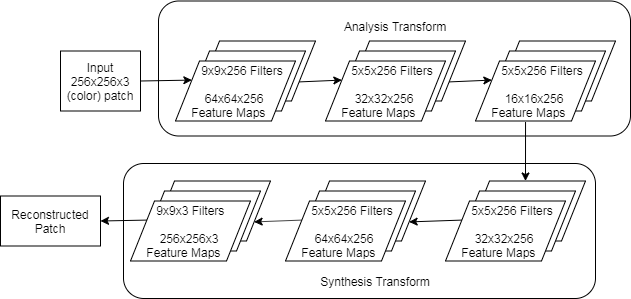}
\caption{Architecture of CNN used as feature extractor.}
\label{fig:cnn_arch}
\end{figure}

We use the local normalization non-linearity inspired by local gain control behaviors in biological visual systems in the 256-channel end-to-end compression architecture \cite{Balle17a}. The CNN architecture used in the proposed method is shown in Fig. \ref{fig:cnn_arch}. The analysis transform block progressively down-samples the input image patch by a factor of 4,2,2 respectively, and uses the forward Generalized Divisive Normalization activation \cite{Balle17a} for all except the last layer. The synthesis transform block progressively up-samples the output of the analysis transform block by a factor of 2,2,4 respectively, and uses the \textit{inverse} Generalized Divisive Normalization activation \cite{Balle17a} for all except the last layer. However, contrary to the authors in \cite{Balle17a}, our aim is not image compression; Hence, we remove the rate-distortion term from the loss function and re-train the network on the DIV2K dataset \cite{div2k1,div2k2}. Random, overlapping $256\times256$ random patches are extracted from each 2K resolution color image. All patches are aggregated and shuffled into batches of 32 patches. Thus, the training is completely unsupervised, and does not use any subjective scores. Note that the CNN in Fig. \ref{fig:cnn_arch} is used for feature extraction, but not score prediction, thereby making our proposed method opinion-\textit{unaware}.

\subsection{Non-parametric modelling of feature distribution}
We use Kernel Density Estimation with Epanechnikov kernels to model any arbitrary-shaped distribution of the features in the encoded layer of the auto-encoder architecture (output of analysis transform block), for building the natural model. This choice was motivated by experiments which showed that the distribution of those features do not conform well to any well-known distribution. This is unlike other NR-IQA methods which mostly fit different types of parametric distributions to hand-crafted features for training and testing. The benefit of using a non-parametric approach is that we need very little information about the underlying distribution. In such scenarios, we cannot properly specify a parametric model. Thus, we can think of non-parametric models as much ``broader'' than parametric ones. Specifically, the kernel density estimator model usually just assumes that the probability density function of the \textit{true} distribution from which the data are sampled satisfies `smoothness' conditions like continuity or differentiability. Eq. \ref{eq:kde} describes a kernel density estimator fitted to $n$ observations $X_{1}$,...,$X_{n}$, where $h$ (positive, chosen empirically) is the \textit{bandwidth}, and $K$ is the function representing the kernel, such that it outputs only positive values which sum (integrate) to 1 over the set of all observations (real numbers).

\begin{equation}
\label{eq:kde}
\hat{f}_{n}(x)=\frac{1}{n h} \sum_{i=1}^{n} K\left(\frac{X_{i}-x}{h}\right)
\end{equation}

\subsection{Opinion- and Distortion-unaware training and score prediction}
After opinion-unaware training of the natural model, we predict the score for any given input image by comparing its distribution of encoded features with those of the natural model, using KL-Divergence. Thus, neither the training nor the testing phase uses any subjective scores or prior knowledge of any expected type(s) of distortions, making the proposed method opinion-and-distortion-\textit{unaware}.

\section{Results}
\label{sec:results}

\subsection{Dataset}
Two primary requirements for an IQA method \cite{ma2017waterloo} are monotonicity (monotonic increase in predicted scores with increasing degree of input image distortion), and consistency (consistent ranking of images with the same distortion type and content, but different distortion levels). To evaluate the same for our proposed method and compare against other state-of-the-art methods, we randomly selected 20 images from the General-100 dataset \cite{gen100} and distorted them with five different levels of each of (Gaussian) blurring, (AWGN) noise and (JPEG-2000) compression artefacts. Thus, our test dataset had 20$\times$(1 clean + 15 distorted) = 20$\times$16 = 320 images.

\subsection{Context of comparison with state-of-art NR-IQA methods}
We compare the proposed method with three state-of-art methods that operate under similar as well as relaxed constraints. As explained earlier, the proposed method is both opinion-unaware (OU) and distortion-unaware (DU) and we first compare it against another ``(OU, DU)'' method, NIQE \cite{6353522}. Next, we compare with an opinion-\textit{aware}, distortion-unaware ``(OA, DU)'' method, BRISQUE \cite{6272356}. Lastly, we compare against an opinion-unaware, distortion-\textit{aware} ``(OU, DA)'' method, PIQE \cite{7084843}. Understandably, NIQE has similar constraints as proposed method, whereas BRISQUE and PIQE operate under relaxed constraints. Thus, it is easier for the last two methods to perform better than our proposed method, because they have more information available to them, although their application scenarios are much more limited than the proposed method, as explained earlier. To remind the reader, OA methods like BRISQUE require supervised training on subjective scores which are difficult to obtain and standardize, and suffers from generalization concerns. DA methods like PIQE have unpredictable performance for (combinations of) distortion types they haven not been designed to detect.

\subsection{Comparison metrics}
Pearson (Eq. \ref{eq:pearson}), Kendall (Eq. \ref{eq:kendall}) and Spearman (Eq. \ref{eq:spearman}) correlation of predicted quality scores with distortion levels (0 through 5) were calculated. The average over all images and distortion types for all methods are reported in Table \ref{tab:result}.

Pearson's correlation coefficient $\rho_{P}$ measures the linear relationship between two variables $X$ and $Y$, which have standard deviations $\sigma_x$ and $\sigma_y$ respectively, and co-variance $cov(X,Y)$. $\rho_{P}$ can have a maximum value of $+1$ denoting perfect positive relationship and a minimum value of $-1$ denoting perfect negative relationship between $X$ and $Y$, while a value of $0$ indicates no relationship.

\begin{equation}
\label{eq:pearson}
\rho_{P} = \frac{cov(X,Y)}{\sigma_x \sigma_y}
\end{equation}

Kendall's correlation coefficient $\tau$ quantifies the degree of monotone relationship between two \textit{ranked} variables $X$ and $Y$, each having $n$ observations. Total number of possible pairings of observations from two variables is ${n\choose 2} = \frac{n(n-1)}{2}$. In some of those pairs, the order in which the observations are ranked are same for both variables (``concordant pairs'', $c$) and in other pairs, the order in which the observations are ranked are different for both variables (``discordant pairs'', $d$) such that $n = c+d$ and $S = c-d$ in Eq. \ref{eq:kendall}. It follows that when $c=n$ and $d=0$, $\tau=+1$ (perfect positive correlation), when $c=0$ and $d=n$, $\tau=-1$ (perfect negative correlation), and when $c=d$, $\tau=0$ (no correlation).

\begin{equation}
\label{eq:kendall}
\tau = \frac{c-d}{c+d} = \frac{S}{{n\choose 2}}
= \frac{2S}{n(n-1)}
\end{equation}

Spearman's correlation coefficient $\rho_{S}$ measures the relationship between $n$ observations of two \textit{ranked} variables $X$ and $Y$, where $d_{i}$ is the pairwise difference of the variables' ranks. A value of $+1$ denotes perfect positive correlation, $-1$ indicates perfect negative correlation, and a $0$ value indicates no correlation.

\begin{equation}
\label{eq:spearman}
\rho_{S} = 1- {\frac {6 \sum d_i^2}{n(n^2 - 1)}}
\end{equation}

\begin{table}
\centering
\caption{Performance comparison of proposed method with state-of-art NR-IQA methods operating under similar (NIQE) and relaxed constraints (BRISQUE and PIQE).}\label{tab:result}
\begin{tabular}{|c|c|c|c|c|}
\hline
Method (constraints) & \thead{Pearson\\score} & \thead{Kendall\\score} & \thead{Spearman\\score} & \thead{Time\\(sec.)}\\
\hline
Proposed (OU, DU) &  0.91 & 0.88 & 0.92 & 42.91\\
NIQE \cite{6353522} (OU, DU) &  0.85 & 0.85 & 0.89 & 0.04\\
BRISQUE \cite{6272356} (OA, DU) & 0.78 & 0.68 & 0.75 & 0.04\\
PIQE \cite{7084843} (OU, DA) & 0.89 & 0.90 & 0.94 & 0.06\\
\hline
\end{tabular}
\end{table}

Table \ref{tab:result} shows better correlation scores for proposed method against a state-of-art method operating under similar constraints (NIQE), and even one which operates under relaxed constraints (BRISQUE). Another type of relaxed constraint method (PIQE) performs slightly better for distortions types it has been designed to detect. However, the proposed method's execution time is significantly higher than all compared methods, and thus it has room for improvement.

\section{Conclusion and Future Work}
\label{sec:confut}
 We proposed a no-reference IQA method in an otherwise unexplored intersection of paradigms. We showed how biologically inspired activation in CNN layers can encode image patches in a reduced dimension, that captures the degree of distortion in the patch, without training on subjective opinion scores or assumption about possible distortion types. We showed via objective evaluation, the superior performance of our proposed method over the state-of-art. The next stage of our research will focus on improving the execution time of our proposed method, as well as further validating our proposed method on much larger datasets.

\bibliographystyle{splncs04}
\bibliography{samplepaper}

\end{document}